# 3D distortion-free, reduced field of view diffusion-prepared GRE at 3T

*Sarah McElroy[1,2], Raphael Tomi-Tricot[1,3], Jon Cleary [1,4], Shawna Kinsella[1], Sami Jeljeli[1], Vicky Goh [1,4], Radhouene Neji[1]*

1. School of Biomedical Engineering and Imaging Sciences, King's College London, United Kingdom
2. MR Research Collaborations, Siemens Healthcare Limited, Camberley, United Kingdom
3. Siemens Healthcare, Courbevoie, France
4. Guy's and St Thomas' NHS Foundation Trust

**Corresponding author:**
Dr. Radhouene Neji
School of Biomedical Engineering and Imaging Sciences
King's College London
3rd Floor Lambeth Wing
St Thomas' Hospital
Westminster Bridge Road
London SE1 7EH
United Kingdom
Email: radhouene.neji@kcl.ac.uk

**Word count:** 2971

**Submitted to Magnetic Resonance in Medicine**

**Abstract**

**Purpose:** To develop a 3D distortion-free reduced-FOV diffusion-prepared GRE sequence and demonstrate its in-vivo application for diffusion imaging of the spinal cord in healthy volunteers.

**Methods:** A 3D multi-shot reduced-FOV diffusion-prepared GRE (RFOV-DP-GRE) acquisition is achieved using a slice-selective tip-down pulse in the phase encoding direction in the diffusion preparation, combined with magnitude stabilisers. The efficacy of the developed reduced FOV approach and accuracy of ADC estimates were evaluated in a phantom. In addition, 5 healthy volunteers were enrolled and scanned at 3T using the proposed sequence and a standard spin echo diffusion-weighted single-shot EPI sequence (DW-SS-EPI) for spinal cord imaging. Image quality, perceived SNR and image distortion were assessed by two expert readers and quantitative measurements of apparent SNR were performed.

**Results:** The phantom scan demonstrates the efficacy of the proposed reduced FOV approach. Consistent ADC estimates were measured with RFOV-DP-GRE when compared with DW-SS-EPI. In-vivo, RFOV-DP-GRE demonstrated improved image quality and reduced perceived distortion, while maintaining perceived SNR compared to DW-SS-EPI.

**Conclusion:** 3D Distortion-free diffusion-prepared imaging can be achieved using the proposed sequence

# 1 INTRODUCTION

Diffusion weighted imaging (DWI) is a powerful tool for probing the microstructure to assess a range of pathologies, including stroke, tumours and inflammatory conditions [1,2]. Conventional DWI employs a Stejskal-Tanner pulsed gradient spin echo [3] in combination with a single-shot EPI (SS-EPI) readout, which is efficient and insensitive to motion [4]. However, SS-EPI is prone to distortion, signal loss and blurring, which are exacerbated at higher field strengths due to increased off-resonance, susceptibility gradients and shorter T2* relaxation times. Signal loss and image blurring are also increased with the longer readout durations required for higher spatial resolution, which is a key benefit of imaging at higher field strengths.

Scan acceleration using parallel imaging to reduce the readout duration and the effective echo-spacing can help to reduce but cannot fully eliminate these effects, especially for imaging regions with high susceptibility gradients [5] such as the frontal lobes, skull base, c-spine and near metal implants where shimming of the static magnetic field can be very challenging.

A further reduction in the readout duration and effective echo-spacing can be achieved using multi-shot EPI, including interleaved EPI [6,7] or readout-segmented EPI [8,9]. The main challenge introduced with these multi-shot approaches is the inter-shot non-linear phase errors due to physiological motion in the presence of strong diffusion gradients [10]. It is therefore necessary to correct for these phase offsets before combining the acquired shots. Solutions for phase correction include phase navigators [11,12] or navigator-free approaches to estimate phase variations between shots such as multiplexed sensitivity-encoding MUSE [13].

Another approach to reduce distortions in diffusion imaging is to use reduced field-of-view (RFOV) EPI techniques [14–17] by restricting the FOV in the phase encoding direction using for example outer volume suppression or 2D RF excitation pulses. RFOV approaches offer some advantages over parallel imaging, including simpler phase correction, no g-factor SNR penalty and avoiding artefacts and unwanted signals from short T1, high-signal producing regions (eg. chest and/or back fat for spinal cord imaging) outside of the anatomical region of interest.

To avoid distortion completely, it is necessary to use an alternative imaging readout to EPI. Diffusion preparation (DP) modules have been proposed which can be combined with any imaging readout [18]. The DP module typically consists of a non-selective tip down pulse, followed by diffusion-sensitizing gradients, refocusing pulses and a non-selective tip-up pulse, which stores the diffusion-prepared magnetisation in the longitudinal direction prior to readout. One of the main challenges with diffusion-prepared imaging is that phase errors due to motion and eddy currents will result in spatially variable magnitude errors [19]. Several solutions have been proposed including preparatory gradients [20], motion-compensated diffusion gradients [21], phase-cycling of the tip-up pulse [19,22] and magnitude stabilisers [23,24]. Recent work suggests that magnitude stabilisers are essential for correction of motion-related phase errors [25]. DP has been combined with turbo spin-echo [26–28], balanced-SSFP [16,21,29] and gradient-echo (GRE) readouts [19,22], which each have advantages and limitations which need to be considered for a particular application and field strength. While TSE and bSSFP offer higher relative SNR, GRE has a lower SAR burden, which is advantageous at higher field strengths.

One potential disadvantage of the GRE readout for diffusion-prepared imaging is the introduction of T1 weighting. T1-weighting can be minimised by employing a centric-encoding[22]. T1 effects can also impact on accuracy of calculated ADC values. These biases can be avoided by employing magnitude stabilisers [30], or by employing phase cycling and ensuring the delay time before each diffusion preparation is slightly greater than the T1, which is also the condition for optimal SNR efficiency [19].

In this work we present a novel technique for distortion-free diffusion-prepared imaging with a multi-shot 3D GRE readout, combining centric encoding, magnitude stabilisers and reduced FOV. The technique is demonstrated in a diffusion phantom and for 3T sagittal imaging of the spinal cord, which is particularly prone to distortion using a conventional SS-EPI acquisition. Part of this work has been presented at ISMRM 2024 [31] where a parallel and independent work pursued a similar idea [32].

## 2 METHODS

### 2.1 Sequence design

The sequence (Figure *1*) consists of a twice-refocused diffusion preparation scheme, paired with a centric-encoded 3D gradient-echo readout, followed by a linearly encoded 2D phase correction navigator. Each shot consists of a single $k_z$ partition in the $k_y$-$k_z$ 2D phase encoding plane, where $k_y$ is the in-plane phase encoding direction and $k_z$ is the partition encoding direction, i.e. in each shot all the in-plane phase encoding steps are acquired for one partition encoding step. The diffusion preparation is preceded by a chemical shift selective fat saturation pulse. The tip-down pulse in the diffusion preparation is applied simultaneously with a slab-selective gradient in the phase-encoding direction, thereby restricting the tip-down excitation in the diffusion preparation module to water protons within a reduced FOV. The tip-down excitation is followed by a pair of bipolar diffusion gradients each surrounding a 180° adiabatic hyperbolic secant refocusing pulse. A magnitude stabiliser approach is used to avoid phase-offset-induced magnitude errors and convert these into conventional inter-shot phase errors. Finally, a non-selective 90-degree tip up pulse is applied. It is worth noting that while this acts as a tip up pulse for the inner-volume magnetisation, it acts as a tip down pulse for the unprepared outer volume magnetisation and any recovered fat magnetisation. A spoiler gradient is executed immediately after the diffusion preparation to dephase any remaining transverse magnetisation including outer-volume and/or fat protons. The original phase of the spins is restored by a magnitude stabiliser rewinder gradient, which is played out after each RF readout pulse and importantly acts also as a spoiler gradient for the outer volume and fat magnetisation. The remaining inner-volume inter-shot phase variations are corrected with the 2D phase navigator in the reconstruction: a 2D inverse Fourier transform is applied to each phase navigator followed by a Hamming filter to reduce Gibbs ringing and each imaging shot for each receiver coil. A complex conjugate multiplication is then performed in image space before applying a Fourier transform to produce a phase-corrected signal [11].

## 2.2  Imaging studies

All imaging was performed at 3T (MAGNETOM Biograph mMR, Siemens Healthcare, Erlangen, Germany, VE11P software). Phantom imaging was acquired with a 16-channel head and neck coil and in-vivo imaging was acquired with a 32-channel spine array and posterior section of a 16-channel head and neck coil.

### 2.2.1  Phantom Studies

A diffusion phantom (Caliber MRI, Boulder, CO, USA) with 6 vials containing varying concentrations of PVP and water was scanned.

**Evaluation of Reduced FOV**

The proposed sequence was acquired twice to assess the effectiveness of the RFOV approach: once with a selective and once with a non-selective tip-down pulse. The following parameters were applied: b0 s/mm$^2$ and b 500 s/mm$^2$ with 3 orthogonal diffusion directions, single average, diffusion-preparation duration: 65 ms, TE/TR: 2.0ms/4.4ms, shot TR: 2000ms, FOV: 200mm x 100mm, acquired matrix size: 96 x 48, acquired slice thickness 4mm, number of partitions: 8, slice oversampling = 25%.

**Quantitative Assessment of ADC**

The diffusion phantom was imaged with the proposed 3D RFOV-DP-GRE and a standard 2D DW-SS-EPI sequence to evaluate the accuracy of ADC estimates. The diffusion sequences were acquired with matched parameters where possible, including:

b0 s/mm$^2$ with a single average, b500 s/mm$^2$ with 3 orthogonal diffusion directions, acquired in-plane resolution: 2.1 x 2.1 mm$^2$, slice thickness: 4 mm, number of slices/partitions: 40. Unmatched parameters (DW-SS-EPI vs RFOV-DP-GRE) include: number of averages for b500: 5 vs. 1, FOV: 263 x 430 mm$^2$ vs 200 x 75 mm$^2$, Echo-spacing/TE/FA: 4.20 ms/79 ms/90° vs. 4.21 ms/2.36 ms/10°, Phase-encode lines per shot: 126 vs 41, Acceleration: GRAPPA 2 / None, Bandwidth: 2025 Hz/Px vs. 410 Hz/Px, Fat suppression: Short-Tau Inversion Recovery/Fat-Sat, Acquisition time: 5 min 22 secs vs. 5 mins 12 secs. Additional parameters for RFOV-DP-GRE include: Shot TR: 2000 ms, diffusion preparation time: 65 ms, number of navigator lines: 14, single navigator partition, phase oversampling: 10%, one dummy scan for magnetisation preparation, 4π intra-voxel phase dispersion for the magnitude stabiliser.

ADC maps were calculated for both sequences using mono-exponential fitting of the signal from the b0 and b500 trace-weighted images.

### 2.2.2 In Vivo Evaluation

Five subjects (2 male, 3 female, mean age 34 ± 6 years) were prospectively recruited for the study. This study was approved by the National Research Ethics Service (REMAS 8700) and written informed consent was obtained from all participants for the scan and for inclusion in this study. The thoracic spinal cord was imaged with the proposed 3D RFOV-DP-GRE sequence, a standard 2D DW-SS-EPI sequence and a 2D TSE sequence as anatomical reference. The diffusion sequences were acquired with the same parameters as for the ADC phantom study, with the following exceptions for both sequences: number of slices/partitions: 8 and b 500 s/mm$^2$ images acquired with 5 averages. Additional modifications to the RFOV-DP-GRE acquisition included: FOV: 200 x 50 mm$^2$, phase-encode lines per shot: 27, slice oversampling: 50%, acquisition time: 6 mins 56 secs.

**Qualitative Assessment**

The images were assessed qualitatively in consensus by two expert readers. Overall image quality (1 = poor, 2 = fair, 3 = good, 4 = very good, 5 = excellent), perceived distortion (1 = major distortion resulting in non-diagnostic image quality, 2 = major distortion but not limiting diagnosis, 3 = minor distortion but not limiting diagnosis, 4 = no distortion) and perceived SNR (1 = poor, 2 = fair, 3 = good, 4 = very good, 5 = excellent) were assessed for each subject/acquisition.

**Quantitative Assessment of apparent SNR**

Apparent SNR was measured on b500 images. An ROI segmenting the spinal cord was delineated on the b500 trace image on the imaging slice that captured the centre of the spinal cord. A second ROI was drawn posterior to the cord, in a region with no apparent signal. The apparent SNR was calculated as the mean signal in the spinal cord ROI divided by the standard deviation of the no-signal region. Significance was tested using a paired t-test with a p-value < 0.05 indicating a significant difference.

## 3  RESULTS

### 3.1 Phantom Studies

Images acquired in the diffusion phantom are shown in Figure 2, demonstrating successful suppression of outer volume signal using the RFOV approach described. When the selective tip-down pulse is replaced by a non-selective pulse, signal from the outer volume wraps into the imaged region, resulting in severe artefacts on the resultant image.

ADC maps calculated for the RFOV-DP-GRE sequence and the DW-SS-EPI sequence are shown in Figure 3. The estimated mean and standard deviation are overlaid on each vial.

### 3.2 In Vivo Evaluation

Figure 4 shows example images acquired in a healthy volunteer.

For the DP-3D-GRE sequence, the mean image quality score, perceived distortion and perceived SNR averaged over the two readers was 2.9, 4.0 and 3.0, respectively. For the DW-SS-EPI sequence, the mean scores were 2.1, 2.4, and 2.9. There was no significant difference in the quantitative assessment of apparent SNR between the DP-3D-GRE sequence and the DW-SS-EPI ($18.8 \pm 5.4$ vs. $16.7 \pm 3.7$, $P = 0.45$).

## 4  DISCUSSION

In this study, a 3D RFOV diffusion-prepared GRE sequence is presented for distortion-free diffusion imaging. The proposed sequence demonstrated improved image quality and reduced perceived distortion, while maintaining perceived SNR compared to DW-SS-EPI. Furthermore, all images acquired in-vivo demonstrated effective suppression of outer-volume and fat signal.

The efficacy of the novel RFOV method introduced in this study can be attributed to the outer-volume signal suppression achieved using a slice-selective/non-selective tip-up/tip-down diffusion preparation, combined with a magnitude stabiliser approach, and this combination is the main contribution of this paper. The outer volume is not excited by the tip-down pulse of the diffusion preparation. Therefore, the 'tip-up' pulse and subsequent spoiler gradient serve to saturate the outer volume magnetisation before the readout. On its own, this outer-volume suppression may be insufficient, since small inefficiencies in the outer-volume saturation or regrowth of outer-volume magnetisation over the course of the readout may still significantly contaminate the low-signal diffusion-prepared magnetisation [25]. Therefore, it is important to also recognise the contribution of the magnitude stabiliser rewinder gradient to the suppression of outer volume signal: the dephasing gradient of the magnitude stabiliser in the diffusion preparation, applied before the tip-up pulse, does not influence magnetisation in the outer volume, since it has not been excited by the tip-down pulse, and thus the magnitude stabiliser rewinder gradient serves to spoil any signal from outer-volume magnetisation that recovers during the course of the readout.

Fat signal (which would otherwise contaminate both the image and phase navigator signal) is also successfully suppressed in a similar manner as described above for outer-volume signal, as the fat saturation pulse, when effective, ensures that fat signal is also not excited by the tip-down pulse of the preparation module. While the fat signal was successfully suppressed in all in-vivo images acquired in this study, there may be cases/anatomies where the chemical shift selective saturation fails, especially in regions with high B0 inhomogeneity and magnetic susceptibility gradients. However, this sequence could be combined with a two-echo Dixon readout for fat-water separation to achieve a more robust fat suppression [33].

The RFOV approach is a key element of the sequence design, as this reduces the readout duration and enables acquisition of all the lines of a partition within a single shot, while maintaining sufficient signal for the navigator acquisition which is acquired after the imaging readout and permitting a straightforward phase correction. While alternative navigator-free phase correction approaches exist, such as 'MUSE'[13], the performance of such approaches is degraded for a high number of shots unless more computationally intensive reconstruction techniques are used [26,34].

Compared to diffusion-prepared TSE sequences[27,28,35], RFOV-DP-GRE offers an alternative for non-EPI diffusion imaging with a much lower SAR deposition burden. This could be beneficial especially given that the diffusion preparation requires the use of a pair of SAR-intensive adiabatic refocusing pulses at 3T to achieve robustness to B0 and B1 inhomogeneity effects [36].

Despite the advantages of increased Fourier averages with a 3D acquisition and the lower receiver bandwidths per pixel, the SNR efficiency of the RFOV-DP-GRE sequence in this study is relatively low compared to the DW-SS-EPI sequence. This is in part due to the inherent 50% signal loss incurred by all diffusion sequences which employ magnitude stabilisers due to the creation of a stimulated echo [37]. In addition, the use of a low flip angle GRE readout further reduces the expected signal compared to a spin echo EPI sequence. Two-dimensional diffusion-weighted sequences also benefit from slice interleaving, which increases the scan efficiency compared to the proposed 3D sequence where time is needed for magnetisation recovery after each GRE shot, as evidenced by the significantly shorter scan time of the 2D DW-SS-EPI sequence. However, other factors impacting the SNR efficiency of the RFOV-DP-GRE sequence are application-specific. For example, application of this sequence to regions requiring an increased number of partitions could offer an increase in SNR efficiency, as signal averaging of the high b-value acquisitions could be traded for Fourier averaging, resulting in preserved SNR and scan-time with increased coverage. It is noted that for thicker slabs the 2D navigator correction could be less accurate due to through-slab phase variations[38], and in these cases, it may be necessary to implement an accelerated 3D navigator for robust phase correction[29,35,39–41]. Another application-specific factor impacting SNR is the shot TR. A relatively long shot TR (2 seconds) was used for the RFOV-DP-GRE sequence in this study to ensure sufficient longitudinal magnetisation recovery and a high steady-state signal in the spinal cord. However shorter shot TRs could be achieved and could be beneficial for the study of regions with short T1 relaxation times, resulting in improved SNR efficiency while achieving much lower SAR deposition than for DP-TSE.

A further difference between RFOV-DP-GRE and DP-TSE is that while TSE readouts are affected by T2 decay along the echo train, the proposed RFOV-DP-GRE sequence with a centric trajectory is affected by the T1 decay of the stimulated echoes. While both may lead to blurring artifacts, T1 relaxation times are typically longer than T2 relaxation times and the achieved echo spacings for GRE readouts are usually shorter than for TSE readouts, so it is expected that the blurring effects related to T1 decay are less pronounced than those due to T2 decay.

Pulsatile cord and CSF motion are known to introduce artefacts when imaging the spinal cord [42]. Using multiple averages and diffusion directions as applied for the b500 images in this study helps reduce the effects of motion. While cardiac or pulse triggering have also been used to mitigate against the pulsatility of the spinal cord[43], this is not desirable in the sequence presented as it would introduce a variable magnetisation recovery delay.

This is a proof-of-concept study and thus we limited our exploration of this technique to the thoracic spine in a small number of healthy volunteers. Other applications with restricted imaging regions could also benefit from this RFOV technique, such as the prostate, cervix, lumbosacral/sciatic and optic nerves. Further testing of this sequence is required in regions with increased off-resonance (eg. c-spine, near implants). Our reference for the evaluation of the proposed sequence was a standard clinical diffusion-weighted spin echo EPI sequence, however this is well known to suffer from severe distortions in the spinal cord, and it is expected that other techniques such as readout-segmented EPI or reduced FOV EPI will exhibit significantly more robustness to distortions compared to a standard EPI readout, as shown in previous spinal cord diffusion imaging studies[14,17,44,45].

## 5  CONCLUSION

In this work, we proposed a novel 3D reduced-FOV, diffusion-prepared GRE sequence and showed initial results for in-vivo feasibility of spinal cord diffusion imaging in healthy volunteers. A clinical study in patients is warranted to further evaluate the potential of this technique.

**FIGURE CAPTIONS**

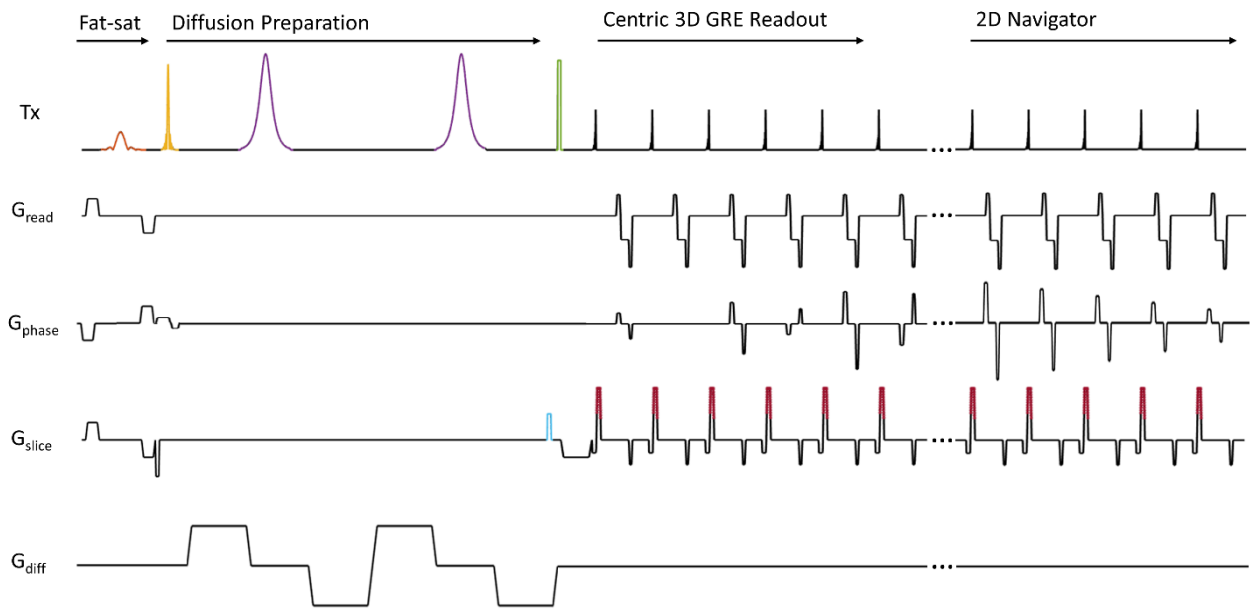

Figure 1: Pulse sequence diagram for 3D diffusion prepared gradient echo acquisition. A chemical shift selective fat saturation pulse (orange) precedes the tip-down pulse (yellow), which is applied in combination with a slab-selective gradient in the phase-encoding direction. This is followed by a pair of bipolar diffusion gradients, each surrounding a 180° adiabatic hyperbolic secant refocusing pulse (purple). A magnitude stabiliser dephasing gradient (blue) is applied before the non-selective tip-up pulse (green). The magnitude stabiliser re-phasing gradient (red) is incorporated in the slice rewinder gradient as an additional moment after each readout excitation pulse.

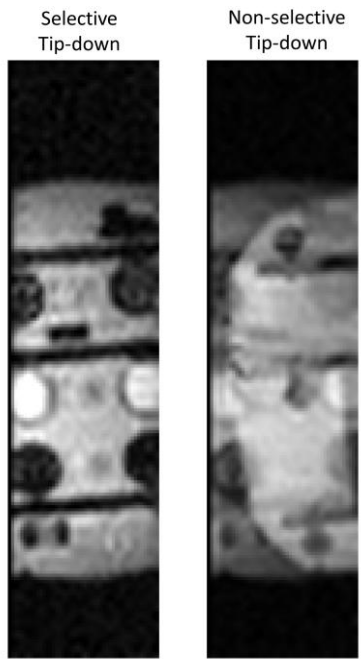

Figure 2: Phantom images acquired with the proposed DP-3D-GRE sequence with a selective (a) and non-selective (b) tip-down pulse. Outer volume signal is not visible when a selective pulse is used, while a non-selective tip-down pulse results in severe wrap artefacts in the imaged region, as a result of signal from the outer volume.

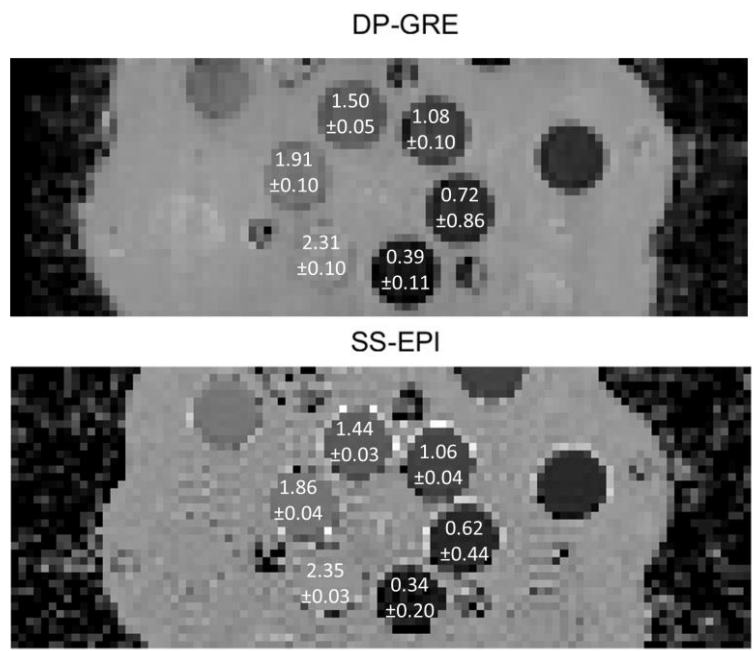

Figure 3: ADC maps calculated from images acquired in the CaliberMRI diffusion phantom with the RFOV-DP-GRE sequence (top) and DW-SS-EPI sequence (bottom). Mean and standard deviation of estimated ADC values are overlaid on each vial (x$10^{-3}$ mm$^2$/s).

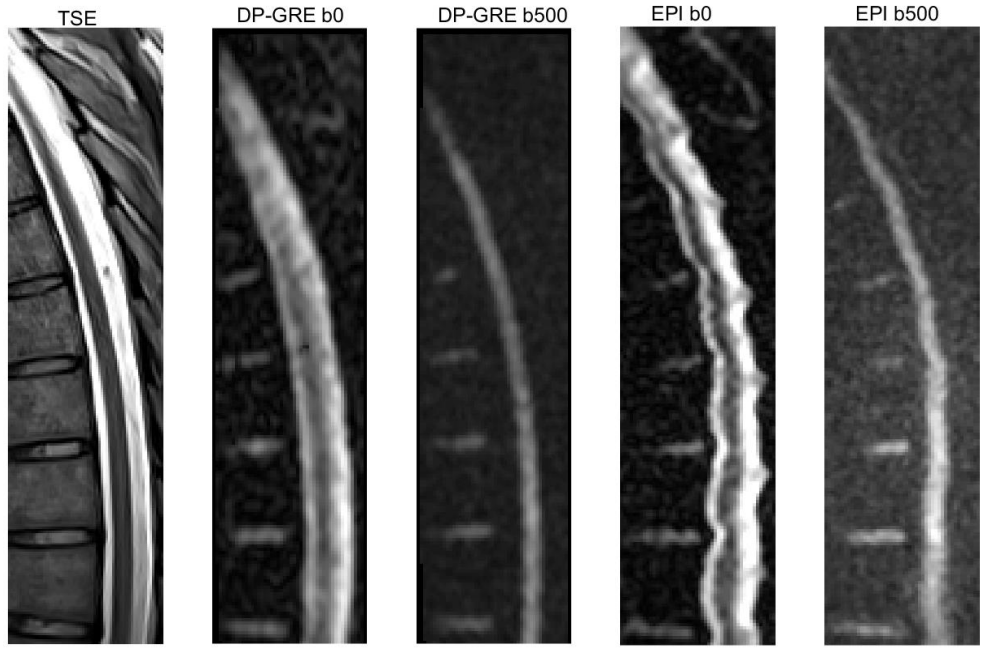

Figure 4: In-vivo example: from left to right: TSE, RFOV-DP-GRE b0, RFOV-DP-GRE b500 Trace-weighted image, DW-SS-EPI b0, DW-SS-EPI b500 Trace-weighted image.